\documentclass [12pt] {article}
\usepackage {amssymb, amsfonts}

\usepackage [cp866] {inputenc}
\inputencoding {cp866}
\begin {document}
\vskip0.3cm \vskip0.5cm  \vskip0.5cm \hfil {\bf {Some notes on
Ishimori's magnet model}} \hfil \vskip0.4cm \hfil {\bf
{E.Sh.Gutshabash}} \vskip0.4cm

\hfil {\small {Institute Research for Physics, Sankt-Petersburg
State University, Russia.}\hfil  \vskip0.2cm \hfil {\small
{e-mail: gutshab@EG2097.spb.edu}}\hfil \vskip0.5cm

\hfil {\bf {Abstract}}\hfil \vskip0.3cm Gauge transformation
properties for an associated linear system of model Ishimori's
magnet model have been discussed. Explicit formulas for the gauge
transformation matrix have been obtained. Darboux Transformation
has been suggested and appropriate dressing relations have been
found.

\vskip0.7cm 1. Ishimori's magnet model is a natural generalization
of the well-known one-dimensional integrable model of Heisenberg's
magnet in two-dimensional case. It has the following form [1]:
$$
{ \bf S} _ t = {\bf S} \wedge ({\bf S} _ {xx} + \alpha ^ 2 {\bf S}
_ {yy}) + u _ y {\bf S} _ x + u _ x {\bf S} _ y, \eqno (1a)
$$
$$
u _ {xx} -\alpha ^ 2u _ {yy} = -2\alpha ^ 2 {\bf S} ({\bf S} _ x
\wedge {\bf S} _ y). \eqno (1b)
$$

Here $ {\bf {S}} = {\bf S} (x, y, t) = (S _ 1, S _ 2, S _ 3) $ is
a three-dimensional vector of magnetization, $ | {\bf S} | = 1,
\:u = u (x, y, t) $ is an auxiliary real-valued field, parameter $
\alpha ^ 2 $ takes values $ \pm 1 $. The case $ \alpha ^ 2 = 1 $
will be denoted a model of Ishimori's magnet-I (MI-I), the case $
\alpha ^ 2 = -1 $ will be denoted the model MI-II.

From the equation (1b) at $ \alpha = 1, \:\xi = (x + y) /2, \:\eta
= (x-y) /2 $ it follows:

$$
u _ {\xi} = \int _ {-\infty} ^ {\eta} {\bf S} ({\bf S} _ {\xi}
\wedge {\bf S} _ {\eta ^ {\prime}}) d\eta ^ {\prime} + u _ 1 (\xi,
t),
$$
$$
\eqno (1c)
$$
$$
u _ {\eta} = \int _ {-\infty} ^ {\xi} {\bf S} ({\bf S} _ {\xi ^
{\prime}} \wedge { \bf S} _ {\eta}) d\xi ^ {\prime} + u _ 2 (\eta,
t),
$$
where $u_1(\xi, t), \:u_2(\eta, t)$ are arbitrary functions, by
virtue of the relations

$$u_1 (\xi, t)=\lim_{\eta \to -\infty}u_{\xi}, \:\:\:\:\: u_2 (\eta,
t)=\lim_{\xi \to -\infty}u_{\eta},$$ they playing a part of
boundary conditions for the model MI-I.

It should be pointed two characteristic particular cases of the
system (1a) - (1b): a). If $ {\bf S} $ does not depend on a
variable $ y $, and $ u = {\mathrm const} $, we obtain an
one-dimensional integrable model of Heisenberg's magnet [2]; b).
If $ {\bf S} $ does not depend on $ t $ (a static limit), $ \alpha
^ 2 = 1 $, $ u = {\mathrm const} $, we deal with so-called
nonlinear $ O (3) $ - sigma-model (an elliptical version) that was
solved in [3-4] by the Inverse Scattering Transform method (using
boundary conditions of the spiral structures type).

An important feature of the model (1) is a presence of so-called
topological charge:

$$
Q_T = \frac {1} {4\pi} \int_{-\infty}^{\infty}
\int_{-\infty}^{\infty} dx\:dy\:{\bf S} ({\bf S}_x\wedge {\bf
S}_y). \eqno (2)
$$
The value remains in evolving the system and represents a mapping
of a unitary 2-sphere onto a 2-sphere: $ \tilde S^2\rightarrow
\tilde S^2$. Such mapping is known to be characterized by a
homotopic group $\pi_2 (\tilde S^2)=Z$, where $Z$ is a group of
whole numbers, which means that a value $Q_T$ should be an
integer. According to the equation (1b) a scalar function $u = u
(x,t)$ has a significance of a production density of a topological
charge.

There are many papers devoted to the model (1). Here we quote just
works [5-7], where to solve the system (1) it was used a number of
methods like the method of Riemann-Hilbert's problem (at $ \alpha
=1$) and the $ \bar {\partial}$-problem method (at $\alpha = i$)
and as well as [8], where a gauge equivalence of two integrable
(2+1)-dimensional systems, namely, the model MI-II and the
Devey-Stewartson's-II, was established for the first time, and, at
last, a paper [9] devoted to analyzing the model (1) from the
point of view of compatible boundary conditions.

The goal of the given work is to apply Darboux's Transformation
method (DT) to the model (1). It will allow to obtain a number of
new additional symmetries of the problem and to study some useful
and important relations characterizing (1). It should be noted
that for the first time the method was applied to (1) in a work
[10], where a rather non-standard approach to DT was employed and
a particular physical problem was solved though. It has to do with
the fact that applying a standard DT method to the corresponding
associated linear problem, in view of its structure, does not
result to the desirable answer ("dressing" \enskip formulas).
Therefore here, with the purpose of a consecutive application of a
standard form of DT, based on the results of [8], at first the
gauge transformation for an initial associated linear system will
be used and only after a matrix of the transformation has been
defined explicitly, Darboux's covariance of the gauged system will
be checked up.

2. To illustrate distinctions between behavior of systems MI-I and
MI-II at the beginning we shall consider the simplest, but rater
important from a physical point of view, special case of waves of
constant amplitude.

Let $\alpha^2=1$ (model MI-I). On putting $S_{+}=S_1+iS_2, \;
S_{-}={\bar S_{+}}, \; S_3={\mathrm const},\: S_{+}=\sqrt {1-S _
3^2} e^{i(kx+ly-\omega_0 t+\phi)}$, where $\:k, \:l, \:\omega_0,
\: \phi \in \mathbb {R}$ are parameters, and substituting this
solution to equations sequent from (1a):

$$
S_{+t}=iS_3\triangle S_{+}+u_yS_{+y}+u_xS_{+x}, \:\:\:\:S_{-}
\triangle S_{+}=S_{+} \triangle S_{-},
$$
considering (1c) and in the assumption $u_1=u_2 \equiv 0$ one
obtains:

$$
\omega_0=S_3(k^2+l^2) .\eqno (3)
$$
Let $ \alpha^2 =-1$ (model MI-II). In this case one has:

$$
S_{+t} =iS_3 (S_{+xx}-S_{+yy})+u_yS_{+y}+u_xS_{+x}, \:\:\: \:S_ 1
(S_{2xx}-S_{2yy})=S_2(S_{1xx}-S_{1yy}).
$$
Using the same anzats (with the assumption, that $S_3={\mathrm
const}) $ and the same boundary conditions, one obtains:

$$
\omega_0=S_3 (k^2-l^2) .\eqno (4)
$$
The discrepancy between signatures of square forms of dispersion
relations (3) and (4) indicates a distinction of behaviour of
appropriate linearized solutions and, hence, as well as the
nonlinear equations depending on a sign of $\alpha^2$. The
formulas (3), (4) generalize a corresponding formula for an
one-dimensional model of Heisenberg's magnet [11] to the
two-dimensional case.

3. Model (1) is a compatibility condition ($\Psi_{yt}=\Psi_{ty}$)
of the next overdetermined linear matrix system:

$$
\Psi_y=\frac {1}{\alpha} S \Psi_x, \eqno (5a)
$$
$$
\Psi_t=-2iS \Psi_{xx}+Q \Psi_x, \eqno (5b)
$$
where $Q=u_yI+\alpha^3u_xS+i\alpha S_yS-iS_x, \:\Psi=\Psi (x, y,
t) \in Mat (2, \mathbb {C}), \:S=\sum_{i=1}^3S_i\sigma_i, \:
\sigma_i$ are Pauli's standard matrixes, $I$ is an unitary $
2\times 2$ matrix. By virtue of its definition the matrix $S$ has
the following properties: $S=S^{\ast}, \; S^2=I, \; {\mathrm
det}\:S=-1$ (a symbol ($\ast$) denotes hermitian conjugation).

One puts

$$
\Psi =\Omega\Phi, \eqno (6)
$$
where $\Phi=\Phi(x, y, t), \:\Omega=\Omega(x, y, t) \in Mat (2,
\mathbb {C})$ and in (5) makes a gauge transformation with the
matrix $\Omega$. Then we have ($U\equiv S$):

$$
\Phi_y=A\Phi_x+B\Phi, \eqno (7)
$$
$$
\Phi_t=L\Phi_{xx}+M\Phi_x+R\Phi, \eqno (8)
$$
where

$$ A=A (x, y, t)=(1/\alpha) \Omega^{-1} U\Omega, \:\:\:B=B (x, y, t) =
(1/\alpha) \Omega^{-1} U\Omega_x-\Omega^{-1} \Omega_y, $$
$$ L=L (x, y, t)=-2i\Omega^{-1} U\Omega, \:\:\:M=M (x, y, t)=
-4i\Omega^{-1} U\Omega_x+\Omega^{-1} Q\Omega, $$
$$R=R(x, y, t)=-2i\Omega^{-1} U\Omega_{xx}+
\Omega^{-1} Q\Omega_x-\Omega^{-1} \Omega_t, $$ $ A, \:B, \:L, \:M,
\:R \in Mat (2, \mathbb {C}),\: L=-2i\alpha A $, and, besides $
A^2=(1/\alpha^2) I, \; L^2=-4I $.

A requirement of the compatibility of (7)-(8) results to relations
including the initial potential and the matrix $\Omega $:

$$
\bigl [A, L \bigr]=0, \eqno (9)
$$
$$
AL_x-L_y-2LA_x+\bigl [A, M \bigr]+\bigl [B, L \bigl]=0, \eqno (10)
$$
$$
A_t+\bigl [A, M_x \bigr]+\bigl [A, R\bigr]+\bigl [B, M\bigr ]
=LA_{xx}+2LA_x+2LB_x+M_y, \eqno (11)
$$
$$
AR_x+B_t+\bigl [B, R]=R_y+LB_{xx}+MB_x. \eqno (12)
$$
Hereinafter we shall restrict ourselves to considering the case $
A \equiv C={\mathrm const}$. Then from (10) it follows
($L_x=L_y=0): M= -2i\alpha B$, and we obtain a closed linear
partial differential equation on the matrix $\Omega$:

$$
2i\Omega_x+2i\alpha U\Omega_y=UQ\Omega .\eqno (13a)
$$
Using the expression for matrix $A$, we have the evolutionary
equation on $\Omega$:

$$
U\Omega_t-\alpha\Omega_tC+U_t\Omega=0.\eqno(13b)
$$

Let $\alpha=1$ (the case MI-I). Then equation (13a) reduces to the
form:

$$
2i(I+U)\Omega_{\xi}=(I+U)Q\Omega .\eqno (14)
$$
Here $\det (I+U)=0$, and characteristic coordinates already
introduced in (1c) are used.

From the relation $\bar Q(-U, -u)=-\sigma_2Q (U, u) \sigma_2$ it
follows an involution:

$$
\Omega (U, u)=\sigma_2 {\bar \Omega}(-U, -u)\sigma_2, \eqno (15)
$$
then the matrix $\Omega$ is represented as:

$$
\Omega (U, u)=\left (\matrix {\omega_{11} (U, u) & - {\bar
\omega}_{21} (-U, -u) \cr \omega_{21} (U, u) & {\bar \omega}_{11}
(-U, -u) \cr} \right). \eqno (16)
$$

To solve the equation (13a) one assumes that

$$
S \to \sigma_3, \:\:\:\:u \to 0, \eqno (17)
$$
as well as that at $r=\sqrt {x^2+y^2} \to \infty$ these limits can
be rapidly achieved.

For further discussions it is useful to introduce a stereographic
projection (here and below we omit a dependence on $t$):

$$
w=w(\xi, \eta)=\frac {S_{+}}{1-S_3} .\eqno (18)
$$
Integrating (14) on the assumption that $\Omega (\xi, \eta) \to I,
\: u(\xi, \eta) \to 0$ at $\sqrt {\xi^2+\eta^2} \to \infty$, on
fulfilling a number of calculations one finds ($\Omega =
\{\omega_{ij} \}, \:i, j = 1, 2$):

$$
\omega_{11}(\xi, \eta)=e^{\sigma_{01}+i\Phi_{01}}, \:\:\: \omega _
{21} (\xi, \eta)=\frac {\omega_{11} (\xi, \eta)} {\bar w}, \eqno
(19)
$$
where

$$
\sigma_{01}=\sigma_{01}(\xi, \eta)=\int_{-\infty}^{\xi}
 d\xi^{\prime} \frac {{\bar w}_{\xi^{\prime}}}
{\bar w(|w|^2+1)}, \:\: \Phi_{01}=\Phi_{01}(\xi, \eta)=-\frac {1}
{2}u(\xi, \eta). \eqno (20)
$$
Thus, the requirement of coincidence of asymptotics of the
function $\omega_{11} (\xi, \eta)$ at $\xi \to \pm \infty$ results
in restriction (bound) for all $\eta \in {\mathbb R}$:

$$
\int_{-\infty}^{+\infty}
 d \xi^{\prime} \frac {{\bar w}_{\xi^{\prime}}}
{\bar w(|w|^2+1)}=0.\eqno (21)
$$
The expression for $\sigma_{01}$ can be rewritten in a more
symmetrical form:

$$
\sigma_{01}=\sigma_{01}(\xi, \eta) =\int_{-\infty}^{+\infty} \int
_ {-\infty}^{+\infty}
 d\xi^{\prime} d\eta^{\prime} \theta (\xi-\xi^{\prime})
\delta (\eta-\eta^{\prime}) \frac {{\bar w}_{\xi^{\prime}}} {\bar
w(|w|^2+1)}, \eqno (22)
$$
where $\theta (.)$ is Heaviside's function.

In the issue the matrix $\Omega$ is written as a product:

$$
\Omega=\Omega^{(11)} \Omega^{(12)}=\left ( \matrix {e^{\sigma _
{01}} & -\frac {1}{w} e^{\bar {\sigma}_{01}} \cr \frac {1} {\bar
w} e^{\sigma_{01}} & e^{\bar {\sigma}_{01}} \cr} \right) \left
(\matrix {e^{i\Phi_{01}} &0 \cr 0 & e^{-i\Phi_{01}} \cr} \right)
.\eqno (23)
$$
It is not difficult to show that if $u (\xi, \eta) \not \to 0$ at
$\sqrt {\xi^2+\eta^2} \to \infty$ and also, generally speaking,
$\Omega \not \to I$, then from the requirement of coincidence of
asymptotics at $\xi \to \pm \infty$ is followed by the necessity
of satisfying the equality (21) and a condition $\Phi_{01}
(-\infty, \eta)=\Phi_{01}(+\infty, \eta)$ for all $\eta \in
{\mathbb R}$.

On using the equation (1b) and rewriting the expression for the
density of a topological charge in terms of variable $w$, one
represents function $u(\xi, \eta)$:

 $$
u(\xi, \eta)=C_1\int \int d\xi^{\prime} d\eta^{\prime} \: G_1
(\frac {\xi-\xi^{\prime}}{2}, \frac {\eta-
 \eta^{\prime}} {2}) \frac {w_{{\eta}^{\prime}}
{\bar w}_{{\xi}^{\prime}}-w_{{\xi}^{\prime}} {\bar w}_{{\eta}^
{\prime}}}{(|w|^2+1)^2}, \eqno (24)
 $$
where $G_1(\xi, \eta)=(1/2) \bigr [\theta (\xi)-\theta (\eta)
\bigl]$ is Green's function for the wave equation, $C_1$ is a pure
imaginary constant, and, hence, the matrix $\Omega$ depends only
on a variable $w$.

Similarly at $\alpha=i$ (the case MI-II), putting $z=x+iy, \;
{\bar z=x-iy}$, from (13a) one obtains an equation:

$$
4i\left (I+U\right) \Omega_{\bar z}=\left (I+U\right) Q\Omega,
\eqno (25)
$$
so in virtue of an involution

$$
\Omega (U, u)=\sigma_2 {\bar \Omega (U, u)} \sigma_2, \eqno (26)
$$
that follows from the symmetry relation $Q(U, u)=\sigma_2 {\bar
Q}(U, u)\sigma_2$, the matrix $\Omega$ takes the form:
$$
\Omega (U, u)=\left (\matrix {e^{\sigma_{02} (z, \bar z)} & -\frac
{1} {w} e^{{\bar {\sigma}_{02}} (z, \bar z)} \cr \frac {1} {\bar
w} e^{\sigma_{02} (z, \bar z)} &e^{{\bar \sigma}_{02} (z, \bar z)}
\cr} \right) e^{\Phi_{02}}, \eqno (27)
$$
where

$$
\sigma_{02} (z, \bar z)=\ln \omega_{11}=\int \int \frac {
d\zeta\wedge {d\bar \zeta}} {2\pi i} \frac {{\bar w}_{\bar \zeta}
(\zeta, \bar \zeta) {\bar w}^{-1} (\zeta, \bar \zeta)} {(z-\zeta)
(1+|w(\zeta, \bar {\zeta})|^2)},
$$
$$
\eqno (28)
$$
$$
\Phi_{02}=\Phi_{02}(z, \bar z)=-\frac {1}{2} u (z, \bar z).
$$
The relation (27) is obtained assuming $\Omega \to I, u (r) \to 0
$ at $r \to \infty$, where $r=|z|=\sqrt {x^2+y^2}$, and applying
the $\bar \partial$ problem technique (see for example [7], [12])
in integrating the equation (25).

Analogously to the case MI-I, from (27) it is possible to exclude
the function $u(z, \bar z)$, on using (1b):

$$
u(z, \bar z)=C_2\int \int \frac {dz^{\prime} \wedge d {\bar
z^{\prime}}} {2i} G_2 (z-z^{\prime}, {\bar z}-{\bar z}^ {\prime})
\frac {w_{{\bar z}^{\prime}} {\bar w}_{z^{\prime}}-{\bar w}_{{\bar
z}^{\prime}} w_{z^{\prime}}} {(|w|^2+1)^2}, \eqno (29)
$$
where $G_2(z, \bar z)=(1/\pi) \ln |z|$ is Green's function of
Laplace's operator, $C_2$ is a real constant, so the matrix
$\Omega $ depends only on a variable $w$.

The expression for $\Omega^{-1}$ for the system MI-II, as has
already been noticed, for the first time was obtained by other
means in [8]. It does not contain the second co-factor in (27),
which is connected with the fact that in [8], as distinct from the
given paper, an associated linear system for a gauge equivalent
counterpart (equation of Davey-Stewartson-II) is assumed to be
known. Moreover, from the relation $S=\Omega \sigma_3 \Omega^
{-1}$, used in [8] in obtaining the gauge equivalence, it follows
that the matrix $\Omega$ can be determined to multiplication by a
diagonal matrix from the right.

Thus, for the system MI-II the relations (27)-(28) generalize the
result obtained in [8] whereas for the systems MI-I corresponding
expressions (23), (19)-(24) are completely new. A gauge
equivalence of the MI-I system and Devey-Stewartson-I one is
absent (as it was shown in [7], it holds only for effectively
one-dimensional models),which means that properties of their
solutions, strictly speaking, can be quite different.

4. Now let us proceed to checking-up the Darboux's covariance of
the system (7)-(8). Let us assume

$$
{\tilde \Phi}=\Phi_x-\tau _ 1\Phi\:\:,\:\:\tau_1=\Phi_{1x} \Phi_1^
{-1},   \eqno (30)
$$
where $\Phi_1$ is some fixed solution of the system at $S=
S^{(1)}$, and $S^{(1)}$ is a fixed solution of (1). Then the
requirement of covariance of (7)-(8) with respect to the
transformation (30) results in the following dressing formulas:

$$
{\tilde A}=A, \:\: \tilde B=B+\bigl [A, \tau_1 \bigr], \:\:\tilde
B =\tau_1B \tau_1+(-B_x+\tau_{1y}-A\tau_{1x}) \tau_1^{-1}, \eqno
(31)
$$

$$
{\tilde M}=M+\bigl [L, \tau_1 \bigr], \:\:\tilde L=L, \:\:\tilde
R=R+M_x+2\tilde L \tau_{1x}+\tilde M\tau_1-\tau_1M. \eqno (32)
$$

It should be noticed that in applying DT (30) the matrix $\Omega$
($\Omega=\Omega (U)$) is also transformed: $\Omega \to \tilde
\Omega $, where $\tilde \Omega \in Mat (2, {\mathbb C}), \:\tilde
\Omega=\tilde \Omega (\tilde U)$. It means that (31)-(32), along
with $\tilde U$ and the matrix $\Omega$ has been calculated, all
include an unknown matrix $\tilde \Omega$ as well, therefore
obtaining an information about new solutions of the model (1) from
this system meets with difficulties.

However, it is possible to act as follows. Let us assume that

$$
\tilde \Omega=K \Omega, \eqno (33)
$$
where $K=K (x, y, t) \in Mat (2, \mathbb {C})$, then from the
former of the relations (31) we immediately find:

$$
{\tilde U}=KUK^{-1}. \eqno (34)
$$
Comparing this expression with a formula of potential
reconstruction obtained in a soliton sector of the problem for the
case of a (one-dimensional) model of Heisenberg's magnet [2], it
is not difficult to see that the matrix $K$ plays a part of
Blyashke's multiplier and it satisfies to generalized unitarity
conditions (see below).

Using now expressions for $M$ and $\tilde M, \; \tilde Q$ by means
of matrixes $\Omega$ and $\tilde \Omega$ correspondingly, and
calculating traces of matrixes $\tilde M$ and $\tilde Q$,
considering (33)-(34) from the former from equalities (32) we find
{\footnote {A more explicit expression for $\tilde u$ will be
given below.}}

$$
\tilde u_y (x, y, t)=u_y+2i\: {\mathrm Sp} \:\left (UK^{-1} K_x
\right). \eqno (35a)
$$
Relations (34), (35a) are a solution of the system (1), provided
that matrix $K$ has been found.

Let us proceed to constructing dressing formulas in terms of the
matrix $K$ for the topological charge (2) assuming that in
applying DT $Q_T \to \tilde Q_T$. For this purpose, taking into
account the identity ${\bf S} ({\bf S}_x \wedge {\bf S}_y)=(1 /
(2i)){\mathrm Sp}\: (SS_xS_y)$ (2) is rewritten as:

$$
Q_T=\frac {1} {8\pi i} \int_{-\infty}^{\infty}
\int_{-\infty}^{\infty} dx dy\: {\mathrm Sp} \left (SS_xS_y
\right). \eqno (36)
$$
Using (33)-(34), for $\tilde Q_T$ we have:

$$
\tilde Q_T=\frac {1} {4\pi i} \int_{-\infty}^{\infty}
\int_{-\infty}^{\infty} dx dy\: {\mathrm Sp} \left \{K^ {-1}K_x
[U, K^{-1} K_y] \right \}. \eqno (37a)
$$

Let us find out some properties of the matrix $K$.

From the dressing relation (31) for $\tilde B$ and using (33)-(34)
we have an equation:

$$
UK^{-1} K_x-\alpha K^{-1} K_y=\alpha \Omega [A, \tau_1]
\Omega^{-1},
$$
or, considering an expression for the matrix $A$:

$$
UK^{-1} K_x-\alpha K^{-1} K_y =[U, \:\Omega \tau_1\Omega^{-1}].
\eqno (38a)
$$
On multiplying both sides of this equation from the right of the
matrix $K^{-1}$ we obtain a closed linear equation on the matrix $
\hat K \equiv K^{-1}$:

$$
U {\hat K}_x-\alpha {\hat K}_y=[U, \:\Omega \tau_1\Omega^{-1}]
{\hat K}. \eqno (38b)
$$
Taking into account the definition of $\tau_1$ and the formula (6)
we have:

$$
U {\hat K}_x-\alpha {\hat K}_y=-[U, \: \Psi_{1x} \Psi_1^{-1}
-\Omega_x\Omega^{-1}] {\hat K}, \eqno (39a)
$$
where $\Psi_1$ is some fixed solution of (5).

Using (13b), (33), we can find the second (evolutionary) linear
equation on matrix $K$:

$$
K_t(U\Omega-\alpha\Omega
C)+K(U_t\Omega-U\Omega_t-\alpha\Omega_tC)=0. \eqno(39b)
$$

The equation (38a) allows rewriting the formula (35a) after
integration over $ y $ in the following form:

$$
{\tilde u}(x, y, t)=u(x, y, t)+2i\alpha \ln {\mathrm det} K+f(x,
t), \eqno (35b)
$$
where it is supposed, that $f(x, t)$ is a real-valued function
that can be defined considering boundary conditions of the type
(1c) and an asymptotic of the matrix $K$ at $y \to \pm \infty$. It
follows that a requirement of realty $\tilde u$ in the case of the
model MI-I results in the restriction: $\ln\: |{\mathrm det}
K|=0$, and, hence, then we have

$$
{\tilde u}(x, y, t)=u(x, y, t)-2 {\mathrm arg} \: {\mathrm det}
K+f(x, t). \eqno (35c)
$$
Using an easily checked identity
$$
{\mathrm Sp} \: \bigl [(K_yK^{-1})_y-\alpha^2 (K_xK^{-1})_x \bigr
]=-\frac {1}{\alpha} {\mathrm Sp} \: \bigl [U (K^{-1})_xK_y-U
(K^{-1})_yK_x \bigr], \eqno (40)
$$
we find:

$$
\tilde Q_T=\frac {i\alpha} {4\pi} \int_{-\infty}^{\infty}
\int_{-\infty}^{\infty} dx dy
 [(\partial_{yy}-\alpha^2\partial_{xx}
) \ln \det K]. \eqno (37b)
$$
The realty of this value for the case MI-I follows from the remark made
above.

It is necessary to note that relations (34), (35) and (37) are in
accord with similar relations deduced by another technique in
[5]-[7], namely, by the $\bar {\partial}$-dressing method and the
Inverse Scattering Transform one.

Having denoted $\Phi[0]=\Phi, \:\Phi[1]={\tilde \Phi}, \:\Phi
[2]={\tilde {\tilde {\Phi}}}, \ldots , \Omega[0]=\Omega, \: \Omega
[1]={\tilde \Omega}, \:\Omega [2]={\tilde {\tilde {\Omega}}},
\ldots, \:K_0=K, \:\Omega[i+1]=K_i\Omega [i], \:i= 0,1, \ldots $,
at next applications of DT we have chains of the form: $\Phi[0]
\to \Phi[1] \to \Phi[2] \ldots, \: \Omega[0] \to \Omega[1]
\to\Omega[2] \ldots, \:K_0 \to K_1 \to K_2 \ldots $.

Thus, after applying a $N$-multiple recurrence of the dressing
procedure, from relations (34), (35), (37) it is easily to write
formulas describing $N$-soliton solutions ($U[N] \equiv S[N]$)
{\footnote {It should be especially emphasized that as distinct
from [5]-[7] in the given paper a spectral problem was not
solved.}}:

$$
U[N]=(\prod_{j=0}^N K_{N-j})U(\prod_{j=0}^N K_{N-j})^{-1}, \eqno
(41)
$$
$$
u[N]=u+2i\alpha \ln (\prod_{j=0}^N {\mathrm det} K_j)+\sum_{i=
0}^N f [i], \eqno (42)
$$
$$
Q_T[N]=\frac {i\alpha} {4\pi} \int_{-\infty}^{\infty}
\int_{-\infty}^{\infty} dx dy
 [(\partial_{yy}-\alpha^2\partial_{xx}
) \ln {\mathrm det} K_N]. \eqno (43)
$$

The other equation for the matrix $K$ can be found substituting
the anzats (34) in (1a). For simplicity we restrict ourselves by
an initial solution {\footnote {Otherwise such an equation would
be too bulky.}} $S=\sigma_3$ taking into account the identity: $(K
\sigma_3K^{-1})_x=\bigl [K_xK^{-1}, K \sigma_3 K^{-1} \bigl]$ and
the formula (35b), as a result we find:

$$
K_tK^{-1}+\frac {1}{2i} \left.\left \{\bigl (K_{xx}+
\alpha^2K_{yy} \bigr) \sigma_3K^{-1}+K \sigma_3\bigl [(K^{-1})_
{xx}+\alpha^2 (K^{-1})_{yy} \bigr] \right.\right \}+
$$
$$
+\frac {1}{i} \bigl [K_x\sigma_3 (K^{-1})_x+\alpha^2K_y\sigma_3
(K^{-1})_y \bigr]-\eqno (44)
$$
$$
-(u+2i\alpha \ln \det K)_yK_xK^{-1}-(u+2i\alpha\ln \det K)_
xK_yK^{-1}=0.
$$

In closing this section it should be noticed that, thus, we have
obtained two, generally speaking, independent techniques to solve the
system (1).

The first one is reduced to finding the matrix $\Omega$ on putting
some initial solution of the system (5) and solving the equation
(13), then from (39) one obtains the matrix ${\hat K}$ (or $K$)
and, in according to the dressing relations constructed above one
has new solutions of (1).

The second one is connected with an immediate solution of the
equation (44) (though in the general case it can be rather
difficult, but it might be possible in elementary special cases)
and with a following employment of the dressing relations.

5. Let $\alpha=1$. Then on using characteristic variables from
(39a) it follows the linear equation:

$$
(I-U){\hat K}_{\xi}+(I+U){\hat K}_{\eta}=-\bigl [U, (\Psi_
{1\xi}+\Psi_{1\eta})\Psi_1^{-1}-(\Omega_{\xi}+\Omega_{\eta})
\Omega^{-1} \bigr] \hat K .\eqno (45)
$$
Besides from (15) one has:
$$
K(U, u)=\sigma_2 {\bar K}(-U, -u)\sigma_2, \eqno (46)
$$
and the same relation for the matrix $\hat K $.

From (45) it is not difficult to show that equations for the
matrix $\hat K$ columns can be reduced to a hyperbolic system of
equations.

Considering (34) and (46) one finds implicit representations for
components $\tilde S \equiv \tilde U$ in terms of matrix elements
of the matrix $K$ and for an arbitrary initial solution $S$:

$$
\tilde S_3=\frac {k_{11}(S, u) {\bar k}_{11}(-S, -u)-k_{21}(S, u)
{\bar k}_{21}(-S, -u)} {\det K}S_3-
$$
$$
-\frac {{\bar k}_{21}(-S, -u){\bar k}_{11}(-S, -u)S_{+}+k_{11}(S,
u)K_{21}(S, u)S_{-}} {\det K},
$$
$$
\eqno (47)
$$
$$
\tilde S_{+}=\frac {2k_{21}(S, u){\bar k}_{11}(-U, -u)S_3+{\bar
K}_{11}^2(-S, -u)S_{+}-k_{21}^2(S, u)S_{-}}{\det K},
$$
where $\det K=k_{11}(S, u) {\bar k}_{11}(-S, -u)+k_{21}(S, u)
{\bar K}_{21}(-S, -u)$, and, as it has already been shown, the
condition $|{\mathrm det} K|=1$ should hold true.

Let us also cite an equation of the form (44) that formally can be
rewritten as a system of equations ($u_{\xi}=u_{\eta}=0$):

$$
K_t+\frac {1}{4i} (K_{\xi \xi}+K_{\eta \eta}) \sigma_3+\frac {1}
{4i} K\sigma_3 (\hat K_{\xi \xi}+\hat K_{\eta \eta}) K+\frac {1}
{2i} K_{\xi} \sigma_3\hat K_{\xi} K+
$$
$$
+\frac {1}{2i} K_{\eta} \sigma_3\hat K_{\eta} K-i(\ln {\mathrm
det} K)_{\xi} K_{\xi}+i(\ln {\mathrm det K})_{\eta} K_{\eta}=0,
$$
$$
\eqno (48)
$$
$$
\hat K_t-\frac {1}{4i}\sigma_3 (\hat K_{\xi \xi}+\hat K_{\eta
\eta})-\frac {1}{4i} \hat K (K_{\xi \xi}+K_{\eta \eta})\sigma_
3\hat K-\frac {1} {2i} \hat KK_{\xi} \sigma_3 {\hat K}_{\xi} -
$$
$$
-\frac {1}{2i} \hat KK_{\eta} \sigma_3 \hat K_{\eta}-i(\ln
{\mathrm det} K)_{\xi} \hat K_{\xi}+i(\ln {\mathrm det K})_{\eta}
\hat K_{\eta}=0.
$$

Let $\alpha=i$. Then from (39a), using variables $z, {\bar z}$, we
have:

$$
(U+I)\hat K_z+(U-I) \hat K_{\bar z}=-\bigl [U, (\Psi_z+ \Psi_
{\bar z})\Psi^{-1}-(\Omega_z+\Omega_{\bar z})\Omega^{-1} \bigr]
\hat K, \eqno (49)
$$
where for the matrix $\hat K$ there exists a symmetry relation:

$$
\hat K (z, {\bar z}, t)=\sigma_2 {\bar {\hat K}}(z, {\bar z}, t)
\sigma_2 .\eqno (50)
$$
It is obvious that the equation (49) can be reduced to an elliptical
system of equations.

Moreover, from (34) and (50) we obtain dressing relations of the form

$$
\tilde S_3(z, {\bar z}, t)=\frac {(|k_{11}|^2-|k_{21}|^2)S_
3-k_{11}k_{21}S_{-}-{\bar k}_{21}{\bar K}_{11}S_{+}}{\det K} ,
\eqno (51)
$$
$$
\tilde S_{+}(z, {\bar z}, t)=\frac {2 {\bar k}_{11}k_{21}S_3+
{\bar K}_{11}^2S_{+}-k_{21}^2S_{-}}{\det K} , \eqno (52)
$$
where $\det K=|k_{11}|^2+|k_{21}|^2$.

It should be noticed that in this case the equation (44) also can
be formulated as a system of nonlinear equations ($u_z=u_{\bar z}
=0$):

$$
K_t+\frac {1}{i}(K_{zz}+K_{{\bar z}{\bar z}})\sigma_3+\frac {1}
{i}K \sigma_3(\hat K_{zz}+\hat K_{{\bar z} {\bar z}})K-
$$
$$
-2i(K_z \sigma_3 \hat K_z+K_{\bar z} \sigma_3\hat K_{\bar z})K+
4i\bigl [(\ln \det K)_zK_z-(\ln \det K)_{\bar z}K_{\bar z} \bigr ]
=0,
$$
$$
\eqno (53)
$$
$$
\hat K_t-\frac {1}{i} \hat K(K_{zz}+K_{{\bar z} {\bar z}}) \sigma
_3\hat K-\frac {1}{i} \sigma_3(\hat K_{zz}+\hat K_{{\bar z}{\bar
z}})+
$$
$$
+2i\hat K (K_z\sigma_3 \hat K_z+K_{\bar z} \sigma_3\hat K_{\bar
z})-4i\hat K \bigr [(\ln \det k)_zK_z-(\ln \det K)_{\bar z}K_
{\bar z} \bigr]\hat K=0.
$$
Despite of complexity and cumbersomeness of equations (48) and (53) they
can be useful as sources of some solutions of the model (1).

6. Now let us establish a link between the approach developed here
and a spectral one. For that it is necessary to introduce a
spectral parameter into the system (5). In (5a) one puts $\Psi=
\psi\exp [(i/\lambda)xI-I\sigma^{(1)}y/(\alpha \lambda)+
2iUt/\lambda^2)]$, where $\lambda \in \mathbb {C}$ is the spectral
parameter, $\sigma^{(1)}=\sigma^{(1)}(x, y, t), \; \psi=\psi(x, y,
t), \; \sigma^{(1)},\; \psi \in Mat (2, \mathbb {C})$, assuming
$[\sigma^{(1)}, U]=0$. Then the function $\psi$ must obey an
equation:

$$
\psi_y-\frac {i}{\alpha \lambda} \psi \sigma^{(1)}=\frac {1}
{\alpha} U\psi_x+\frac {i}{\alpha \lambda} U\psi. \eqno (54)
$$
Representing $\psi$ as $\psi=\sum_{k=0}^{\infty} \psi_k
\lambda^k$, at $|\lambda| \to 0$ from (54), in particular, one
finds: $-\psi_0\sigma^{(1)}=U \psi_0$.

Similarly for the system (7), putting $\Phi=\varphi \exp [
(i/\lambda) xI-i\sigma^{(1)}y/(\alpha \lambda)+2iUt/\lambda^2]$,
one obtains an equation

$$
\varphi_y-\frac {i}{\alpha \lambda} \varphi\sigma^{(1)}=
A\varphi_x+\frac {i}{\lambda} A\varphi+B\varphi .\eqno (55)
$$
Putting $ \varphi=\sum_{k=0}^{\infty} \varphi_k\lambda^k, |
\lambda| \to 0$, one has: $A=-(1/\alpha) \varphi_0\sigma^{(1)}
\varphi_0^{-1}=(1/\alpha) \Omega^{-1}U \Omega.$ Thus, $
\sigma^{(1)}=-U, \:\psi_0=\Omega \varphi_0$, and from an equality
for the matrix $A$ one finds: $\varphi_0=\Omega^{-1}$, and hence,
$\psi_0=1$.

7. In summary it should be noted that in the given paper the main
accent was made on the specificity of a DT application to the
model (1) without exhibiting explicit solutions because a rather
wide number of the solutions as well as their classification have
been given in [5-7]. On the other hand, a series of integrable
models (Myrzakulov's magnets) that are some modifications (or
generalizations) of Ishimori's model have been proposed in papers
[13]-[15]. But owing to appropriate associated linear systems of
these models are similar, the DT application technique used above
is applicable also to the series though it can result in changing
gauge transformation matrixes. A differential-geometrical
interpretation of a number of quoted relations might be of
interest as well.

The author is thankful to P.P.Kulish for supporting.

\vskip 1cm \hfil {\bf {REFERENCES}} \hfil \vskip 1cm

1. \parbox [t] {12.7cm}
       {{\em Y.Ishimori}, - Progr. Theor. Phys. {\bf 72} (1984), 33.}
\vskip0.3cm 2. \parbox [t] {12.7cm}
       {{\em L.A.Takhtadjan and L.D.Faddeev}, - Hamiltonian approach in the
       theory of solitons. M., Science (1986).}
\vskip0.3cm 3. \parbox [t] {12.7cm}
       {{\em E.Sh.Gutshabash and V.D.Lipovski}, - Teor.Math.Phys {\bf 90}
        No.2 (1992), 175.}
\vskip0.3cm 4. \parbox [t] {12.7cm}
       {{\em G.G.Varzugin, E.Sh.Gutshabash and V.D.Lipovski}, -
       Theor.Math.Phys {\bf 104} No.3 (1995), 513.}
 \vskip0.3cm 5. \parbox [t] {12.7cm}
       {{\em V.G.Dubrovsky and B.G.Konopelchenko}, - Coherent
       structures for the Ishimori equation. 1. Localized solitons
       with the stationary boundaries. Preprint No.90-76. Institute
       of Nuclear Physics. Novosibirsk. 1990.}

\vskip0.3cm 6. \parbox [t] {12.7cm}
       {{\em V.G.Dubrovsky and B.G.Konopelchenko}, - Coherent
       structures for the Ishimori equation. 2. Time-depend boundaries.
       Preprint No. 91-29. Institute
       of Nuclear Physics. Novosibirsk. 1990.}

\vskip0.3cm 7. \parbox [t] {12.7cm}
       {{\em B.G.Konopelchenko}, - Solitons in Multidimensions. World
       Scientific. 1993.}

\vskip0.3cm 8. \parbox [t] {12.7cm}
       {{\em V.D.Lipovski and A.V.Shirokov}, - Funct.An. Appl.,
       {\bf 104}, No.3 (1989), 65.}
\vskip0.3cm 9. \parbox [t] {12.7cm}
       {{\em I.T.Habibullin}, - Theor.Math.Phys, {\bf 91},
       No.3 (1992), 363.}

\vskip0.3cm 10. \parbox [t] {12.7cm}
       {{\em K.Imai and K.Nozaki}, - Progr. Theor. Phys, {\bf 96} (1996),
       521.}

\vskip0.3cm 11. \parbox [t] {12.7cm}
        {{\em H.C.Fogedby}, - J.Phys. A, {\bf 13} (1980), 1467.}

 \vskip0.3cm 12. \parbox [t] {12.7cm}
        {{\em M.J.Ablowitz and P.A.Clarkson}, - Solitons, Nonlinear
        Evolution Equations and Inverse Scattering. Cambridge
        University Press.1991.}

 \vskip0.3cm 13. \parbox [t] {12.7cm}
        {{\em R.Myrzakulov}, - On some integrable and
        nonintegrable soliton equations of magnets. Preprint KSU.
        Alma-Ata. 1987.}

\vskip0.3cm 14. \parbox [t] {12.7cm}
        {{\em G.N.Nugmanova}, - The Myrzakulov equatuions: the
        Gauge equivalent counteparts and soliton solutions.
        Nonintegrable soliton equations of magnets. Preprint KSU.
        Alma-Ata. 1992.}

\vskip0.3cm 15. \parbox [t] {12.7cm}
        {{\em N.K.Bliev, G.Nugmanova, R.N.Syzdukova and R.Myrzakulov},
        - Soliton equations in 2+1 dimensions: reductions,
        bilinearizations and simplest solutions. Preprint CNLP
        No.1997-05. Alma-Ata.1997; solv/int 990214.}

\end {document}